\pdfoutput=1
\documentclass[journal=acsodf, manuscript=article]{achemso}

\usepackage[utf8]{inputenc} 
\usepackage[T1]{fontenc}    
\usepackage[flushleft]{threeparttable}

\usepackage{url}            
\usepackage{booktabs}       
\usepackage{amsfonts}       
\usepackage{nicefrac}       
\usepackage{microtype}      
\usepackage{amsmath}
\usepackage{mathtools}
\usepackage{graphicx}
\usepackage{subcaption}
\usepackage[]{placeins}
\mciteErrorOnUnknownfalse
\graphicspath{ {graphics/} }
\newcommand{\beginsupplement}{%
        \setcounter{table}{0}
        \renewcommand{\thetable}{S\arabic{table}}%
        \setcounter{figure}{0}
        \renewcommand{\thefigure}{S\arabic{figure}}%
     }

\author{Julien Horwood}
\affiliation[InVivo AI]
{InVivo AI}
\alsoaffiliation[Mila]
{Mila, Université de Montréal}
\email{julien@invivoai.com}
\author{Emmanuel Noutahi}
\affiliation[InVivo AI]
{InVivo AI}
\email{emmanuel@invivoai.com}

\title[]{Molecular Design in Synthetically Accessible Chemical Space via Deep Reinforcement Learning}

\begin{document}
\maketitle

\begin{abstract}
The fundamental goal of generative drug design is to propose optimized molecules that meet predefined activity, selectivity, and pharmacokinetic criteria. Despite recent progress, we argue that existing generative methods are limited in their ability to favourably shift the distributions of molecular properties during optimization. We instead propose a novel Reinforcement Learning framework for molecular design in which an agent learns to directly optimize through a space of synthetically-accessible drug-like molecules. This becomes possible by defining transitions in our Markov Decision Process as chemical reactions, and allows us to leverage synthetic routes as an inductive bias. We validate our method by demonstrating that it outperforms existing state-of the art approaches in the optimization of pharmacologically-relevant objectives, while results on multi-objective optimization tasks suggest increased scalability to realistic pharmaceutical design problems.  
\end{abstract}

\section{Introduction}
\label{sec:motivation}
 Following advances in generative modelling for domains such as computer vision and natural language processing, there has been increased interest in applying generative methods to drug discovery. However, such approaches often fail to address numerous technical challenges inherent to molecular design, including accurate molecular reconstruction, efficient exploration of chemical space, and synthetic tractability of generated molecules. Further, these approaches bias the generation of molecules towards the data distribution over which they were trained, restricting their ability to discover truly novel compounds. Previous work \cite{ you_graph_2019, zhou_optimization_2019} has attempted to address these issues by framing molecular design as a reinforcement learning problem \cite{sutton2018reinforcement}, in which an agent learns a mapping from a given molecular state to atoms that can be added to the molecule in a step-wise manner. These approaches generally ensure validity of the generated compounds and avoid the need to learn a latent space mapping from the data. However, they do not address the issue of synthetic tractability, and the proposed atom-by-atom environment transitions prevent rapid exploration of chemical space.

We instead approach the problem in a way that incorporates a favourable bias into the Markov Decision Process. Specifically, we define the environment's state transitions as sequences of chemical reactions, allowing us to address the common issue of synthetic accessibility. While ensuring synthesizability of computationally-generated ligands is challenging, our framework treats synthesizability as a feature rather than as a constraint. Our approach, deemed REACTOR (REACTion-driven Objective Reinforcement), thus addresses a common limitation of existing methods, whereby the synthetic routes for generated molecules are unknown and require challenging retro-synthetic planning. Importantly, the REACTOR framework is able to efficiently explore synthetically-accessible chemical space in a goal-directed manner, while also providing a theoretically-valid synthetic route for each generated compound. 

We benchmark our approach against previous methods, focusing on the task of identifying novel ligands for the D2 dopamine receptor, a G protein-coupled receptor involved in a wide range of neuropsychiatric and neurodegenerative disorders \cite{dopaminereview}. In doing so, we find that our approach outperforms previous state-of-the-art methods, is robust to the addition of multiple optimization criteria, and produces synthetically-accessible, drug-like molecules by design.

\section{Related Work}
\label{sec:related}
Computational drug design has traditionally relied on domain knowledge and heuristic algorithms. Recently, however, several machine learning based generative approaches have also been proposed. Many of these methods, such as ORGAN \cite{guimaraes_objective-reinforced_2018}, take advantage of the SMILES representation using Recurrent Neural Networks (RNNs) but have difficulties generating syntactically valid SMILES. Graph-based approaches \cite{de_cao_molgan_2018, liu_constrained_2019, jin_junction_2019} have also been proposed and generally result in improved chemical validity. These methods learn a mapping from molecular graphs to a high-dimensional latent space from which molecules can be sampled and optimized. In contrast, pure reinforcement learning algorithms such as \cite{ you_graph_2019, zhou_optimization_2019} treat molecular generation as a Markov Decision Process, in which molecules are assembled from basic building blocks such as atoms and fragments. However, a core limitation of existing methods is the forward-synthetic feasibility of proposed designs. To overcome these limitations, \citeauthor{button_automated_2019} \cite{button_automated_2019} propose a hybrid rule-based and machine learning approach in which molecules are assembled from fragments under synthetic accessibility constraints in an iterative single-step process. However, this approach is limited in terms of the flexibility of its optimisation objectives, as it only allows for generation of molecules similar to a given template ligand.

In order to have practical value, methods for computational drug design must also make appropriate tradeoffs between molecular \textit{generation}, which focuses on the construction of novel and valid molecules, and molecular \textit{optimization}, which focuses on the properties of the generated compounds. While prior work has attempted to address these challenges simultaneously, this can lead to sub-optimal results by favouring either the generation or the optimization tasks. Generative models generally do not scale well to complex property optimization problems, as they attempt to bias the generation process towards a given objective within the latent space while simultaneously optimizing over the reconstruction loss. These objectives are often conflicting, making goal-directed optimization difficult and hard to scale when multiple reward signals are required. This is generally the case in drug design, where drug candidates must show activity against a given target as well as favourable selectivity, toxicity, and pharmacokinetic properties.

In contrast, atom-based reinforcement learning addresses the generative problem via combinatorial enumeration of molecular states \cite{zhou_optimization_2019} or \textit{a posteriori} verification of molecules\cite{you_graph_2019}. These solutions are often slow, and create a bottleneck in the environment's state transitions that limits effective optimization. \\

\section{Methodology}
\label{sec: method}

In this work, we decompose generation and optimization by delegating each problem to a distinct component of our computational framework. Specifically, we allow an Environment module to handle the generative process, using known chemistry as a starting point for its design, while an Agent learns to effectively optimize compounds through interactions with this Environment. By disambiguating the responsibilities of each component, and by formalizing the problem as a Markov Decision Processes (MDPs), we allow the modules to work \textit{symbiotically}, exploring chemical space both more efficiently and more effectively.

We begin with a short overview of Markov Decision Processes and Actor-Critic methods for reinforcement learning before defining our framework in detail. 

\subsection{Background}

\subsubsection{Markov Decision Processes}

A Markov Decision Process (MDP) \cite{10.2307/24900506} is a powerful computational framework for sequential decision-making problems. An MDP is defined via the tuple $(\mathcal{S}, \mathcal{A}, \mathcal{R}, \mathcal{P})$, where $\mathcal{S}$ defines the possible states, $\mathcal{A}$ denotes the possible actions that may be taken at any given time, $\mathcal{R}$ denotes the reward distribution of the environment, and  $\mathcal{P}$ defines the dynamics of the environment. Interactions within this framework give rise to trajectories of the form $(s_0,a_0,r_1,s_1,a_1,....,r_T,s_T)$, with $T$ a terminal time step. Crucially, an MDP assumes that: 

\begin{equation}
    p(s_{t+1}, r_{t+1}|a_t,s_t,r_t,...,a_0,s_0) = p(s_{t+1},r_{t+1}|a_t,s_t)
\end{equation}{}

where $t$ denotes discrete time steps.

This definition states that all prior history of a decision trajectory can be encapsulated within the preceding state, allowing an agent operating within an MDP to make decisions based solely on the current state of the environment. This assumption provides the basis for efficient learning, and holds under our proposed framework. An agent's mapping from any given state to action probabilities is termed a \textit{policy}, and the probability of an action $a \in \mathcal{A}$ at state $s$ is denoted $\pi(a|s)$.

\subsubsection{Policy Optimization}

The underlying objective of a Reinforcement Learning agent operating in an MDP is to optimize its policy to maximize the expected return from the environment, until termination at time $T$, defined for any step $t$ by:

\begin{equation}
    E_\pi[G_t] = E_\pi[\sum_{m=t+1}^{T} \gamma^{m-t-1}r_m]
\end{equation}{}{}

where $\gamma$ is a discount factor determining the value of future rewards, and the expectation is taken over the experience induced by the policy's distribution. Several approaches exist for learning a policy that maximizes this quantity. In value-based approaches, Q-values of the form $Q: \mathcal{S} X \mathcal{A} \longrightarrow \mathbb{R}$ are trained to estimate the scalar value of action-value pairs as estimates of the expected return. A policy is then derived from these values through strategies such as $\epsilon$-greedy control \cite{sutton2018reinforcement}. Alternatively, policy-based approaches attempt to parameterize the agent's behaviour directly, for example through a neural network, to produce $\pi_\theta(a|s)$. While our framework is agnostic to the specific algorithm used for learning, we choose to validate our approach with an actor-critic architecture \cite{Konda_Tsitsiklis_2000}. This approach combines the benefits of learning a policy directly using a policy network $\pi_{\theta}$, with a variance-reducing value network $v_{\theta'}$. Specifically, we use a synchronous version of A3C \cite{DBLP:journals/corr/MnihBMGLHSK16}, which is amenable to high parallelization and further gains in training efficiency. The Advantage Actor-Critic (A2C) objective function at time $t$ is given by

\begin{align}
    L(\theta, \theta^{\prime}) &= \log(\pi_{\theta}(a_t|s_t)) \sum_{i=0}^k(r_{t+i} + \gamma^{i}v_{\theta^{\prime}}(s_{t+i}) - v_{\theta^{\prime}}(s)) + \beta\mathcal{H}(\pi_{\theta}(s_t) \\
    &= \log(\pi_{\theta}(a_t|s_t))A_t(s_t,a_t, \theta^{\prime}, k) + \beta\mathcal{H}(\pi_{\theta}(s_t))
    \label{eq: a2c}
\end{align}{}

Intuitively, maximization of equation \ref{eq: a2c}'s first term involves adjusting the policy parameters to align high probability of an action with high expected return, while the second term serves as an entropy regularizer preventing the policy from converging too quickly to sub-optimal deterministic policies. 

\subsection{Molecular Design via Synthesis Trajectories}

A core insight of our framework is that we can embed knowledge about the \textit{dynamics} of chemical transitions 
into a Reinforcement Learning system for guided exploration. In doing so, we induce a bias over the optimization task which, given its close correspondence with natural molecular transitions, should increase learning efficiency while leading to better performance across a larger, pharmacologically relevant chemical subspace. \\

\begin{figure}[!htp]
\centering
    \includegraphics[width=0.9\linewidth]{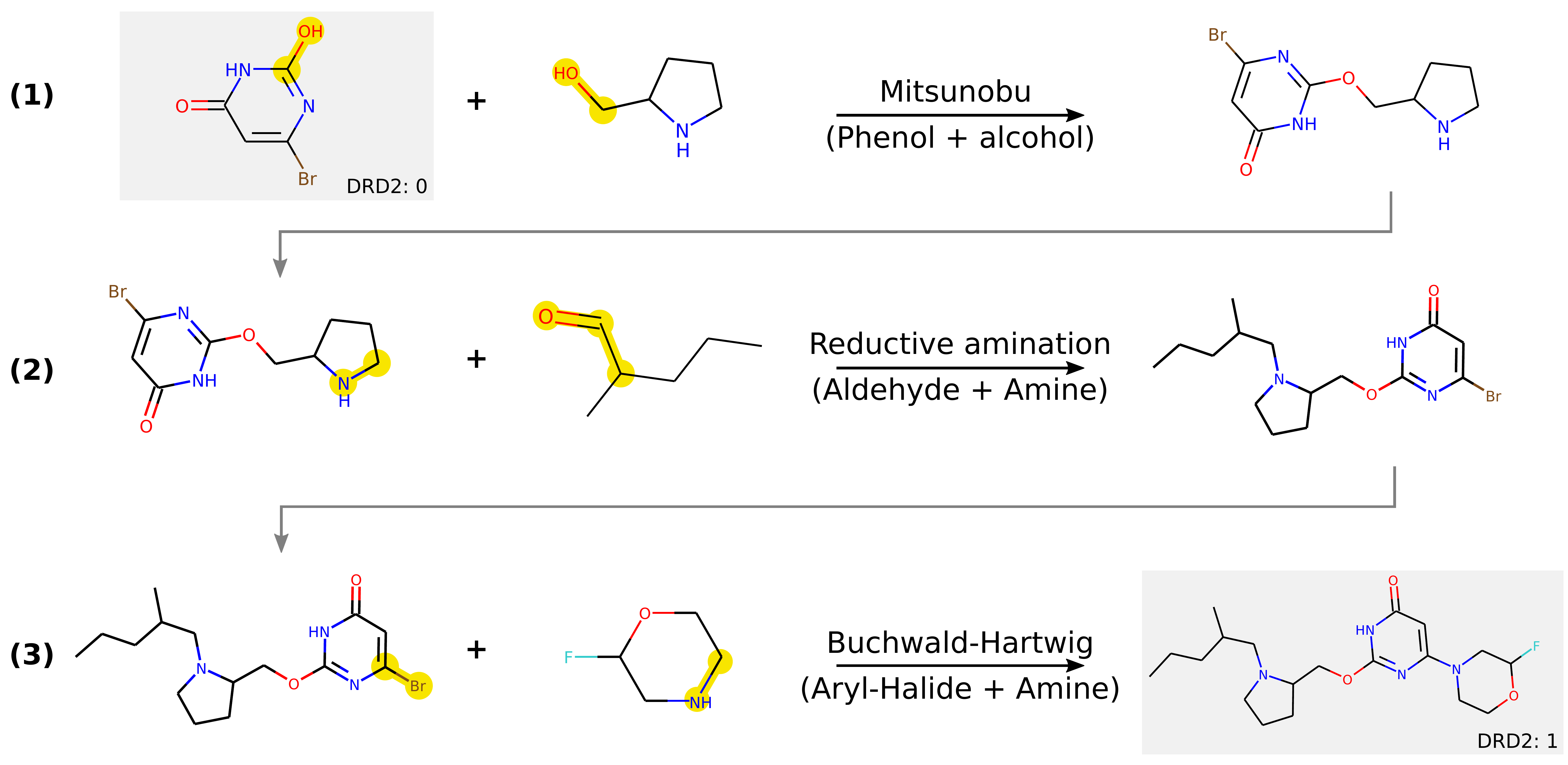}
    \caption{A trajectory taken by the REACTOR agent during the optimization of affinity for the Dopamine receptor D2.This trajectory provides a high-level overview of a possible synthesis route for the proposed molecule in three steps: (1) a Mitsunobu reaction, (2) a reductive amination and (3) a Buchwald–Hartwig amination.  We note that although the proposed route is theoretically feasible, it would not be the first choice for synthesis and can easily be optimized. Nevertheless, it remains an important indication of synthesizability. We also note here that the agent learns a policy that produces structures containing a pyrrolidine/piperidine moiety, which have been shown as actives against dopamine receptors. \cite{martelle2008review,gilligan1992novel}
    }
    \label{fig: trajectories}
\end{figure}

\begin{figure}[!h]
    \centering
    \includegraphics[width=\textwidth]{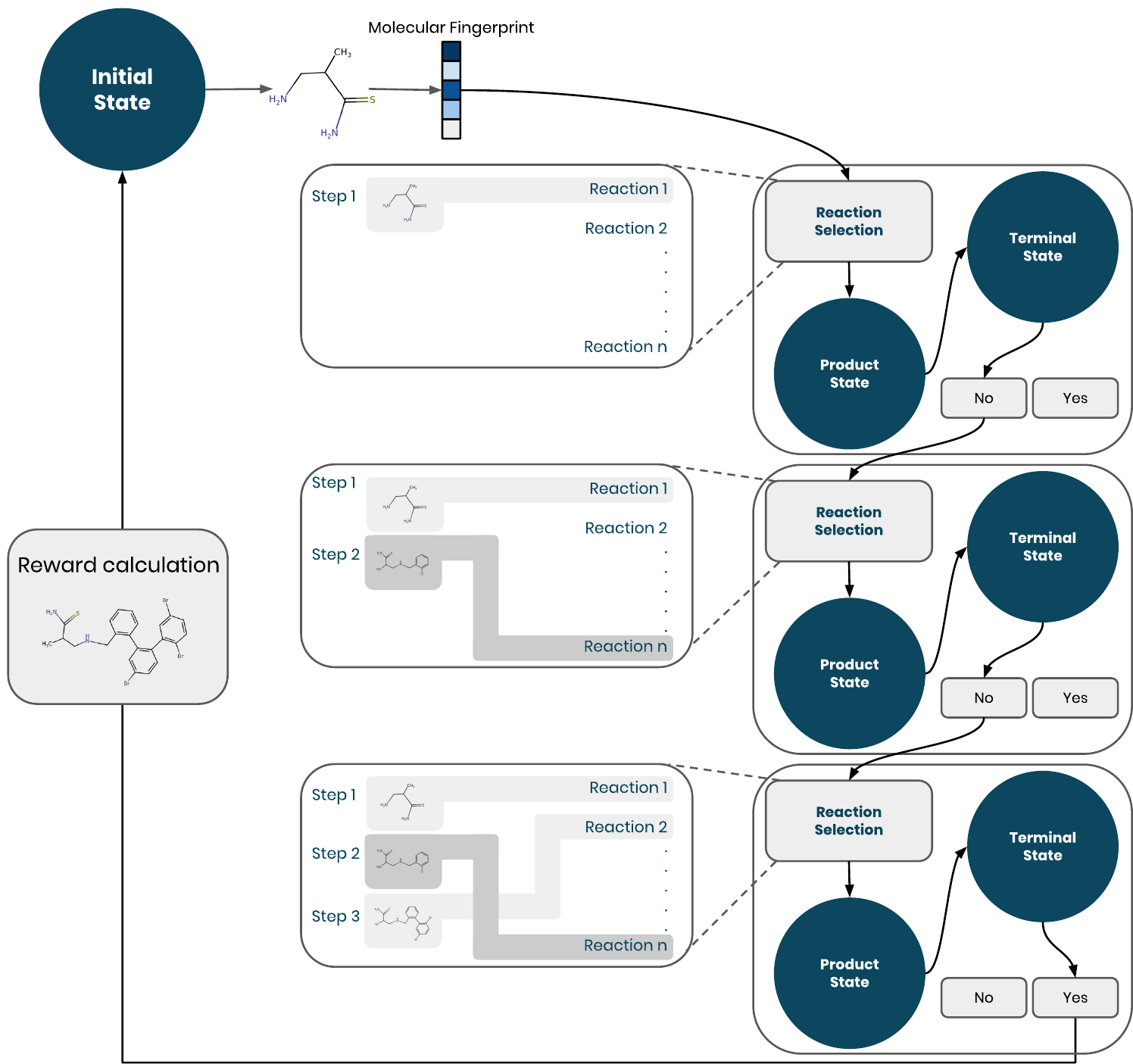}
    \caption{Overview of the REACTOR framework. Each episode is initialized with a molecular building block. At each step, the current
    state is converted to its fingerprint representation, and the policy model selects a reaction to be performed. A reactant selection heuristic
    completes the reaction to generate the next state in the episode, while a reward of 0 is returned. Instead, if the terminal action is selected, the current state is considered as the final molecule and its reward is used to update the policy's parameters.}
    \label{fig:framework}
\end{figure}

We propose embedding this bias into the transition model of an MDP by defining possible transitions as true chemical reactions. In doing so, we gain the additional
benefit of built-in synthetic accessibility, in addition to immediate access to a synthesis route for generated compounds. Lack of synthesizability is a known constraint of prior generative approaches in molecular design \cite{Gao_Coley_2020}. The REACTOR approach addresses this constraint by embedding synthesizability directly into the framework, leveraging synthetic routes as an inductive bias. This is demonstrated in  Figure \ref{fig: trajectories}, where a sample trajectory is provided by REACTOR for a DRD2-optimized molecule, while a high-level overview of our framework is presented in Figure \ref{fig:framework}.

{}

\subsubsection{Framework Definition}

{}

We define each component of our MDP as follows:
\subsubsection*{State Space $\mathcal{S}$}
    We allow for any valid molecule to comprise a state in our MDP. Practically, the state space is defined 
    as $\{f(m) |m \in \mathcal{M}\}$, with $f$ a feature extraction function, $\mathcal{M}$ the space of molecules reachable
    given a set of chemical reactions, initialization molecules, and available reactants. We use Morgan Fingerprints \cite{Rogers_Hahn_2010} with bit-length 2048 and radius 2 to extract feature vectors from molecules. These representations have been shown to provide robust and efficient featurizations, while more computationally-intensive approaches like Graph Neural Networks are yet to demonstrate significant representational benefit \cite{Correction_FP, Kearnes_McCloskey_Berndl_Pande_Riley_2016}.
\subsubsection*{Action Space $\mathcal{A}$}
    In its general formulation, the action space of our framework is defined \textit{hierarchically}, enabling the potential application of novel approaches for hierarchical reinforcement learning. Specifically, we define a set of higher-level actions $\mathcal{A}_o$ as a curated list of chemical reaction templates, taking the form: 
    
    \begin{equation}
    R := r_1 + r_2 + ... + r_k \rightarrow (p_1,...,p_m)    
    \end{equation}

    Each $r_i$ corresponds to a reactant, while each $p_j$ is a product of this reaction. We make use of the SMARTS syntax \cite{SMARTS} to
    represent these objects as regular expressions. We append to the high-level actions a terminal action, allowing the agent to learn to terminate an episode when the current state is deemed optimal for the objective. At step $t$, the state $s_t$ thus corresponds to a \textit{single} reactant 
    in any given reaction. It is necessary to select which molecular blocks should fill in the remaining pieces for a given state and 
    reaction selection. This gives rise to a set of primitive actions $\mathcal{A}_i$ corresponding to a list of reactants derived from 
    the reaction templates, which we also refer to as chemical building blocks. In contrast with previous methods \cite{you_graph_2019,zhou_optimization_2019}, which establish a deterministic 
    start state such as an empty molecule or carbon atom, we initialize our environment with a randomly-sampled building block which matches at minimum one reaction template. This ensures that
    a trajectory can take place and encourages the learned policies to be generalizable across different regions in chemical space. 

    For our experiments, we work with two-reactant reaction templates and select missing reactants based on those which will most improve the next state's reward. We also select the chemical product in this manner when more than one product is generated. Doing so 
    collapses our hierarchical formulation into a standard MDP formulation, with the reaction selection being the only decision point. Additionally, it is likely that for any step $t$, the set of possible reactions is smaller than the full action space. In order to increase both the scalability of our framework (by allowing for larger reaction lists) and the speed of training, we use a mask over unfeasible reactions. This avoids the need for the agent to learn the chemistry of reaction feasibility, and reduces the effective dimension of the action space at each step. We compare policy convergence when using a masked action space to a regular action space formulation in Figure S1. The policy then takes the form $\pi(a_t|s_t, M(s_t,\mathbf{R}))$, with $M$ the environment's masking function and $\mathbf{R}$ the list of reaction templates.

\subsubsection*{Reward Distribution $\mathcal {R}$} 
    Appropriate reward design is crucial given that it drives the policy optimization process. In Graph Convolutional Policy Networks \cite{you_graph_2019},
    intermediate and adversarial rewards are introduced in order to enforce drug-likeness and validity of generated compounds. In MolDQN \cite{zhou_optimization_2019}, these requirements are ignored, and while optimization performance increases, desirable pharmaceutical properties are often lost. In the REACTOR framework, the separation between the agent and the environment allows us to maintain property-focused rewards that guide optimization while ensuring chemical 
    constraints are met via environment design. 

    We use a deterministic reward function based on the property to be optimized. In Table \ref{fig: denovo}, this corresponds to the binary prediction of compounds binding to the D2 Dopamine Receptor (DRD2). In Table S1, these are the penalized calculated octanol-water partition coefficient (cLogP) and quantitative estimate of drug-likeness (QED) \cite{qed}. In order to avoid artificially biasing our agent towards greedy policies, we remove intermediate rewards and provide evaluative feedback only at termination of an episode. While we feel this is a more principled view on the design process, \citeauthor{zhou_optimization_2019} \cite{zhou_optimization_2019} have also suggested that using an intermediate reward discounted by a decreasing function of the step $t$ may improve learning efficiency. 
    We further apply a constraint based on the atom count of a molecule to be consistent with prior work. When molecules exceed the maximum number of atoms (38), the agent observes a reward of zero.
    
    \subsubsection*{Transition Model $\mathcal{P}$} 
    In the template-based REACTOR framework, state transitions are deterministic. We therefore have $p(s_{t+1}|s_t,a_t) = 1$, according
    to our choice of reaction and the subsequent reactant-product selection. When modifying the reactant-selection policy, either via a stochastic 
    heuristic such as an epsilon-greedy reactant selection, or learned hierarchical policies, state transitions over the higher level actions $A_o$ become stochastic according
    to the internal policy's dynamics.

\subsubsection{Building Block Fragmentation}
In order to maximize the exploration capacity of the REACTOR agent, it is desirable to scale up the size of both the reaction template and reactant lists. However, current Reinforcement Learning methodology is poorly suited for very large discrete action spaces. In particular, there are approximately 76000 building blocks available for our experiments, with a wide range of possibilities matching a given reaction template position. While certain approaches propose learning a mapping from continuous to discrete action spaces \cite{chandak_learning_2019, dulac-arnold_deep_2016}, we instead mitigate the dimensionality of the reactant space directly. Indeed, we leverage the BRICS \cite{brics} retrosynthesis rules to reduce our original reactant set to one of approximately 5000 smaller blocks. This reduces the reactant space dimension by an order of magnitude while rendering the transitions in space less extreme, and thus more flexible. Additionally, we may limit the size of the set of reactants under consideration at any given step, treating this as a hyper-parameter. For our experiments, we set this to 100 reactants, finding little variation when selecting reactants in a greedy manner.

\section{Results and Discussion}
\label{sec:experiments}
To validate our framework, we benchmark its performance on goal-directed design tasks, focusing primarily on predicted activity for the D2 Dopamine Receptor. We frame this objective as a sparse reward, using a binary activity indication to simulate a hit discovery setting. In order to maintain consistency with experiments done in prior work, we perform additional experiments on penalized cLogP and QED, with the results presented in Supplementary Material. 

In order to better understand the exploration behaviour of our approach, we also investigate the nature of the trajectories generated by the REACTOR policies, showing that policies retain drug-likeness across all optimization objectives, while also exploring distinct regions of chemical space. 

\subsection{Experimental Setup}
    \textbf{Reaction Data}
    For these experiments, the set of reactions used was obtained from \citeauthor{reactions} \cite{reactions}, with the final list consisting of 127 reactions following curation for specifity and validity. The set of reactants are drawn from PubChem \footnote{A mapping of SMILES to PubChem ID is available upon request} \cite{pubchem}, totalling 76208 building blocks matching the reaction templates. Following the retrosynthesis methodology introduced above, these lists were reduced to approximately 5000 smaller reactants, with 90 reaction templates matching these blocks. This allows us to make the space of action possibilities more tractable, while rendering the exploration of chemical space more flexible due to each transition corresponding to smaller steps in space. Naturally, this action space does not encompass all chemical transformations which may be of interest in a general setting. However, it is straightforward to extend the reaction templates and associated building blocks to tailor the search space to the data available for a given design objective.
    
    \textbf{Empirical Reward Models}
    While generative models are biased by their data distribution, RL-driven molecule design may be biased implicitly by training data used for an empirical reward model. Thus, it is crucial that these models provide robust generalization. A model which is overly simplistic, as is seen for the cLogp experiments, may lead to agents exploiting particular biases, leading to pharmacologically undesirable molecules. Training details for the DRD1, DRD2, DRD3 and Caco-2 models are found in the Supplementary Material.  
    
    \textbf{Baselines}
    We compare our approach to two recent methods in deep generative molecular modelling, JT-VAE and ORGAN \cite{jin_junction_2019, guimaraes_objective-reinforced_2018}. Each of these approaches was first pre-trained for up to 48h on the same compute facility, a single machine with 1 NVIDIA Tesla K80 GPU and 16 CPU cores. Property optimization was then performed using the same procedures as described in the original papers. We also compare our method with two state-of-the-art reinforcement learning approaches, Graph-Convolutional Policy Networks and MolDQN \cite{you_graph_2019, zhou_optimization_2019}. Each algorithm was run using the open-sourced code from the authors, while we enforced the same reward function implementation across methods to ensure consistency. We ran GCPN using 32 CPU cores for approximately 24 hours (against 8 hours in the original paper), and MolDQN for 20000 episodes (against 5000 episodes in the original paper). In addition, we added a steepest-ascent hill-climbing baseline using the REACTOR environment to demonstrate that for simple, mostly greedy objectives such as cLogP and QED, simple search policies may provide reasonable performance.
    In contrast, learned traversals of space become necessary for complex tasks such as DRD2.
    
    \textbf{Evaluation}
    Given the inherent differences between generative and reinforcement learning models, evaluation was adapted to remain consistent within each class of algorithms. JT-VAE and ORGAN were evaluated based on decoded samples from their latent space, using the best results across training checkpoints, with statistics for JT-VAE computed over 3 random seeds. Given the prohibitive cost of training ORGAN, results are given over a single seed and averaged over three sets of 100 samples. Other baselines were compared based on three sets of 100 building blocks used as starting states. Statistics are reported over sets, while the statistics of the initial states are shown by BLOCKS.
    
    We prioritize evaluation of each method based on the total number of active molecules identified, as determined by the environment reward model, given that this corresponds most to the underlying objective of de novo design. Indeed, in a hit discovery scenario, a user may be most interested in identifying the maximal number of unique potential hits, leaving potency optimization to later stages in the lead optimization process. We denote this quantity by ``\textit{Total Actives}'' in Table \ref{fig: denovo}. ``\textit{Mean Activity}'' corresponds to the percentage of generated molecules which are predicted active for the DRD2 receptor. In both Table \ref{fig: denovo} and Table S1, mean reward (``\textit{Mean Activity}'') was computed based on the set of unique molecules generated by each algorithm, in order to avoid artificially favouring methods which often generate the same molecule. Diversity corresponds to the average pairwise Tanimoto distance among generated molecules, while "\textit{Scaff. Similarity}" corresponds to the average pairwise similarity between the scaffolds of the compounds, as implemented by the MOSES repository\cite{moses}. Finally, we limited the number of atoms to 38 for all single-objective tasks, as done in prior work \cite{zhou_optimization_2019, you_graph_2019, jin_junction_2019}, and to 50 for the multi-objective tasks.

\subsection{Goal-Directed De Novo Design}

\begin{table*}[h]
	\centering
    \resizebox{\textwidth}{!}{%
  \begin{threeparttable}
    \setlength{\tabcolsep}{0.2\tabcolsep}
	\caption{Goal-Directed Molecule Design} \label{tab:fragments}

		\begin{tabular}{ccccccc}
			\toprule

			Objective & Method & Total Actives & Mean Activity &Diversity & Scaff. Similarity & Uniqueness\\
			\midrule
			
			DRD2 & BLOCKS & 3 $\pm$ 0 & 0.03 $\pm$ 0 &  0.94 $\pm$ 0 & N/A{*}  & 1.0 $\pm$ 0.0  \\
			\midrule
			&  Hill Climbing & 43.0 $\pm$ 2.94 & 0.43 $\pm$ 0.03 &  0.878 $\pm$ 0.01 &   0.124 $\pm$ 0.0 &  1.0 $\pm$ 0.0\\
			
			& ORGAN  &  5.333 $\pm$ 0.47 &  0.093 $\pm$ 0.01 &  0.86 $\pm$ 0.01 &  0.577 $\pm$ 0.11 &  0.873 $\pm$ 0.01 \\
			
			& JTVAE  &  4.0 $\pm$ 0.82 &  0.014 $\pm$ 0.0 &  0.934 $\pm$ 0.0 &   0.097 $\pm$ 0.0 &  0.976 $\pm$ 0.01 \\
			
			& GCPN  & 0.0 $\pm$ 0.0 &  0.0 $\pm$ 0.0 & 0.906 $\pm$ 0.0 & 0.12 $\pm$ 0.0 &  1.0 $\pm$ 0.0 \\
			
			& MolDQN  &   9.667 $\pm$ 0.47 &  0.816 $\pm$ 0.08 &  0.6 $\pm$ 0.02 &     N/A &  0.12 $\pm$ 0.02\\
			
			& REACTOR &  \textbf{77.0 $\pm$ 4.32} &  0.77 $\pm$ 0.04 &  0.702 $\pm$ 0.02 &  0.133 $\pm$ 0.01 &  1.0 $\pm$ 0.0  \\
			\bottomrule
		\end{tabular}	\label{fig: denovo}

 \begin{tablenotes}
	\small
	\item {*}Computation of Scaffold Similarity requires the presence of a ring system, thus the N/A.
\end{tablenotes}

 \end{threeparttable}}

\end{table*}

Results on the unconstrained design task show that REACTOR identifies the most active molecules for the DRD2 objective. Furthermore, we observe that REACTOR maintains high diversity and uniqueness in addition to robust performance. This a crucial characteristic, as it implies that the agent is able to optimize the space surrounding each starting molecule, without reverting to the same molecule to optimize the scalar reward signal. In Table S1, REACTOR also achieves higher reward on QED, while remaining competitive on penalized cLogP despite the simplistic nature of this objective favouring atom-by-atom transitions. We note that while MolDQN exhibits higher mean activity, this is attributed to the fact that the optimization tends to collapse into generating the same molecule. This explains why the total number of active molecules identified remains low, despite mean activity suggesting a good performance on the task. 

Training efficiency is an important practical consideration while deploying methods for de novo design.
Generative models first require learning a mapping of molecules to the latent space before training for property optimization. During our experiments, this resulted in more than 48h of training time, after which training was stopped. Reinforcement learning methods trained faster, but generally failed to converge within 24 hours. We ran MolDQN for 20000 episodes, taking between 24 and 48 hours, while GCPN was stopped after 24 hours on 32 CPU cores. In contrast, our approach converges within approximately two hours of training on 40 CPU cores for the cLogP and QED objectives, while consuming less memory than GCPN for 32 cores, and terminates under 24 hours for the D2-related tasks. In order to make effective use of parallelization, we leveraged the implementation of A2C provided by the rllib library \cite{rllib}.

\subsection{Synthetic Tractability and Desirability of Optimized Compounds}
\begin{figure}[!h]
\centering
    \begin{subfigure}{.49\linewidth}
    \centering
    \includegraphics[width=0.9\linewidth]{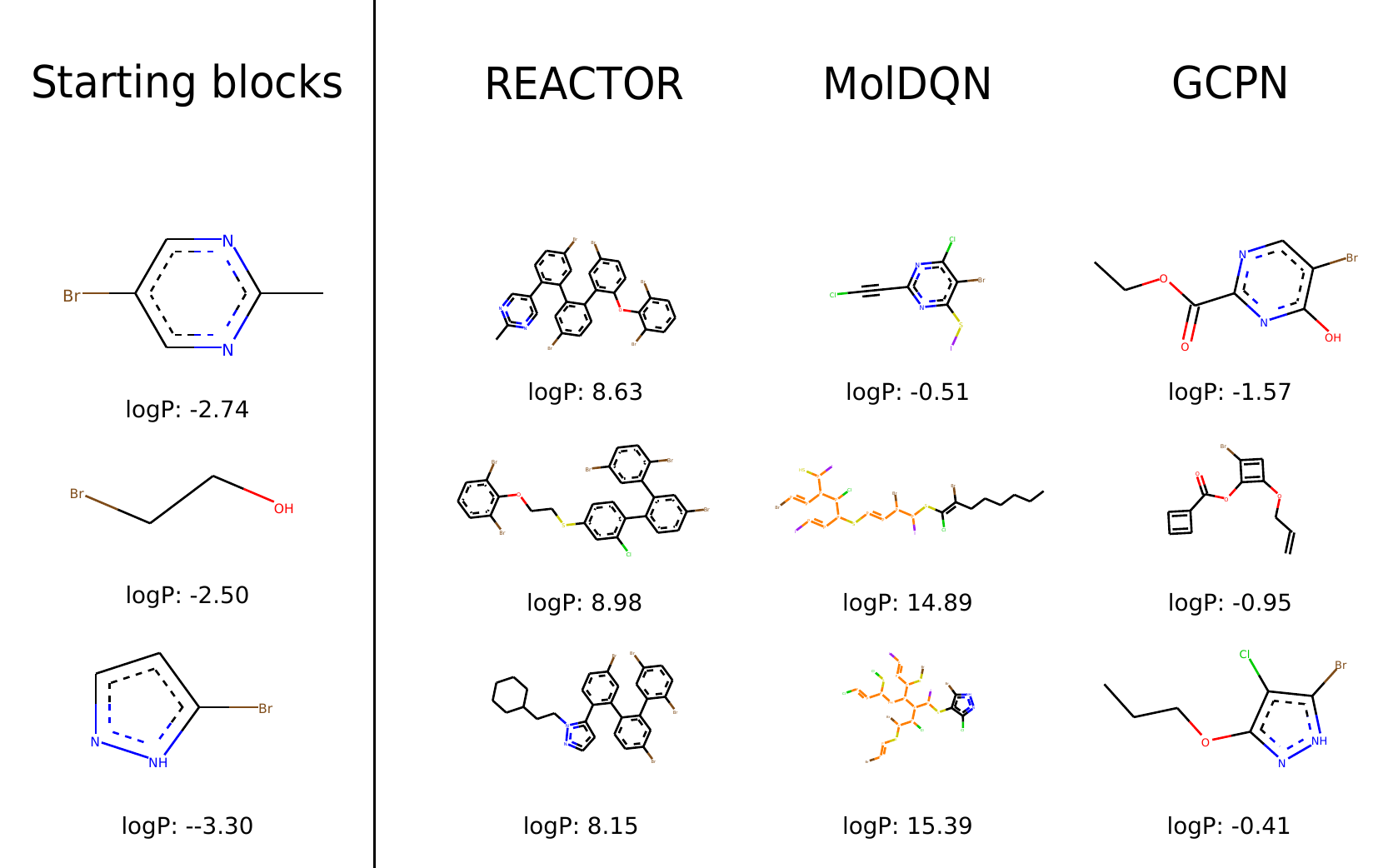}
    \caption{Samples for cLogP objective}
    \end{subfigure}
    \begin{subfigure}{.49\linewidth}
    \centering
    \includegraphics[width=0.9\linewidth]{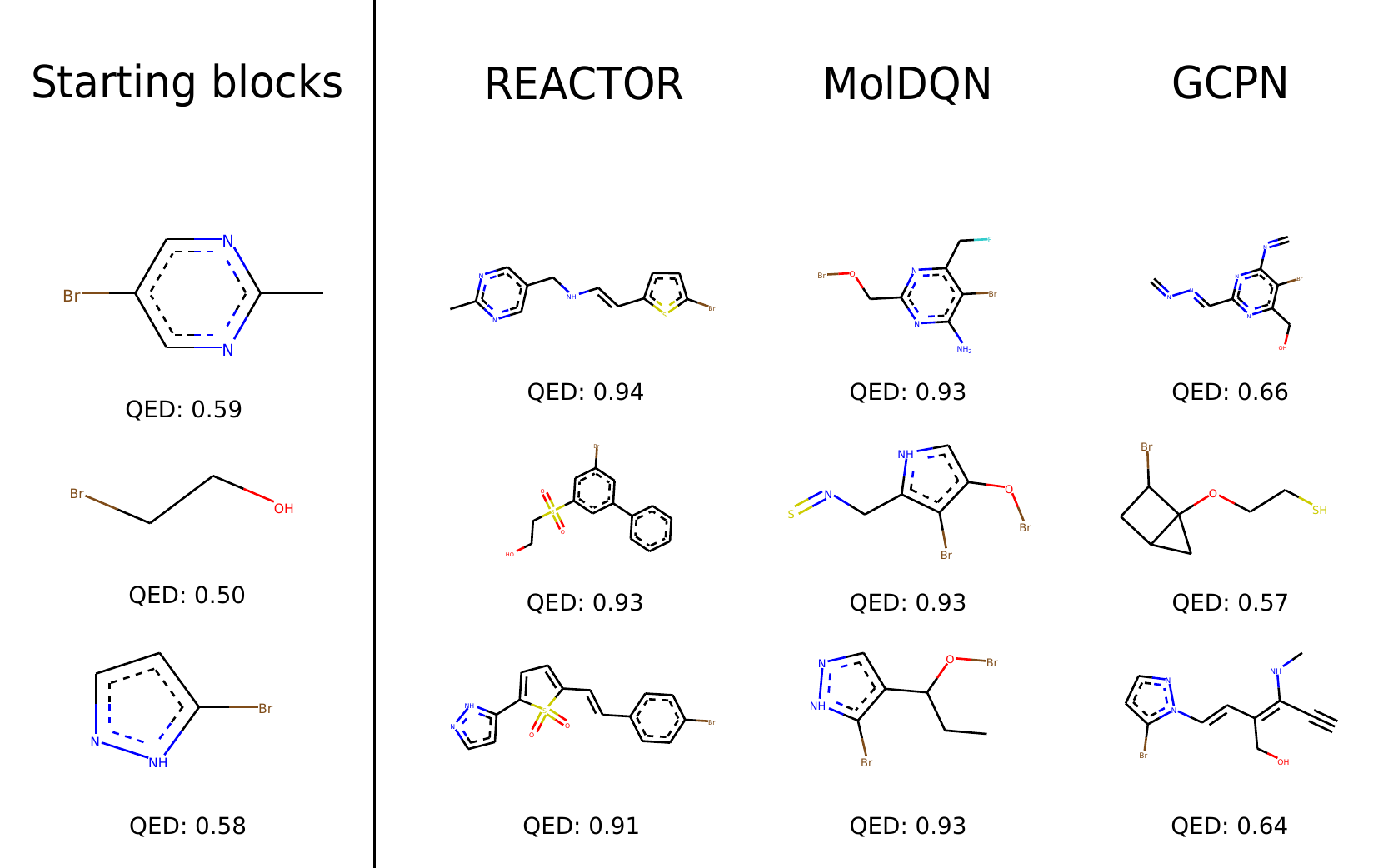}
    \caption{Samples for QED objective}
    \end{subfigure}
    \quad
    \begin{subfigure}{.49\linewidth}
    \centering
    \includegraphics[width=1\linewidth]{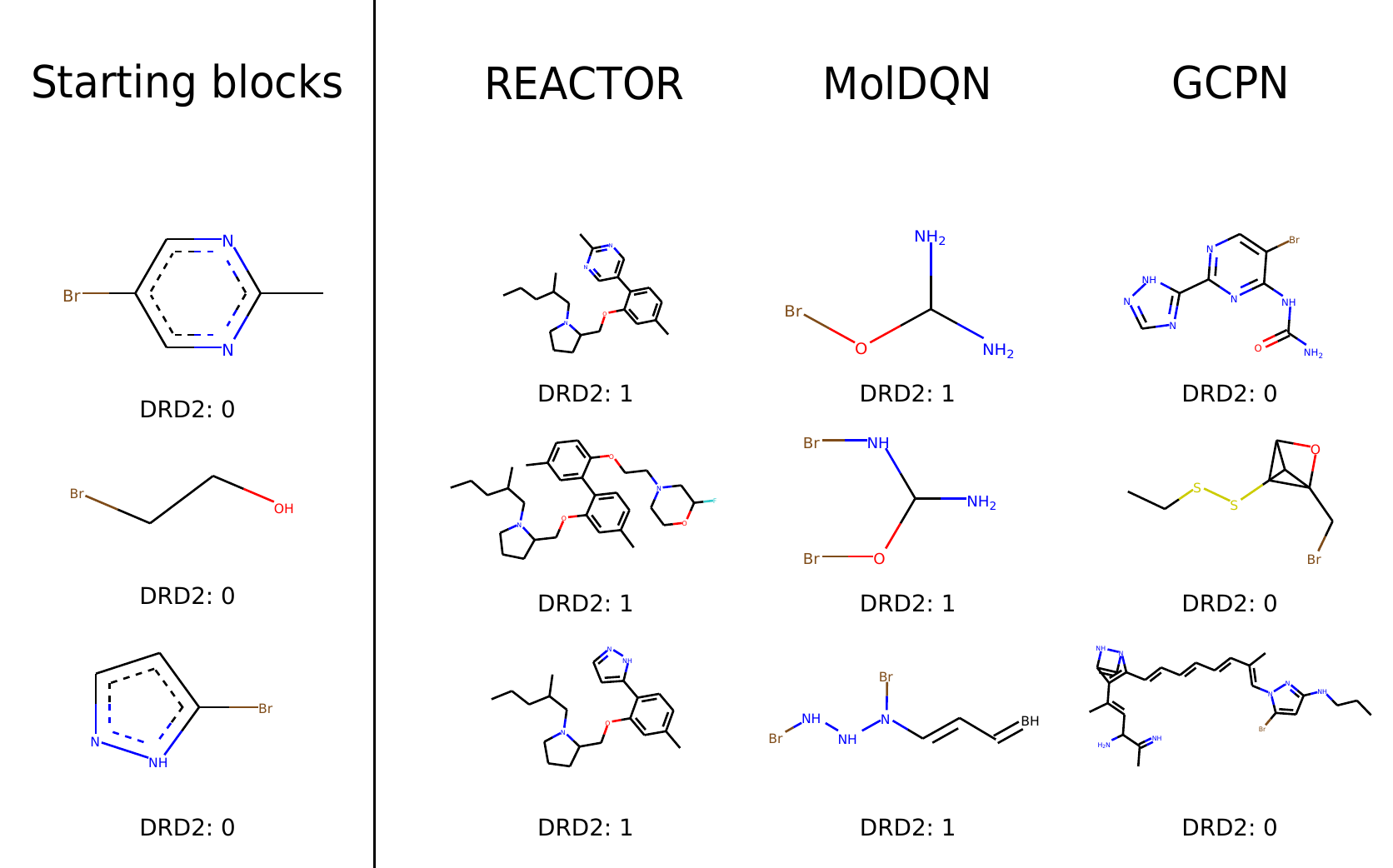}
    \caption{Samples for DRD2 objective}
    \end{subfigure}
\caption{Sample molecules produced for each objective by each RL algorithm}
\label{fig:samples}
\end{figure}

Given the narrow perspective offered by quantitative benchmarks for molecular design models \cite{moses}, it is equally important to qualitatively assess the behaviour of these models by examining generated compounds. Figure \ref{fig:samples} provides samples generated by each RL method across all objectives. Since the computational estimation of cLogP relies on the Wildman-Crippen method~\cite{clogp}, which assigns a high atomic contribution to Halogens and Phosphorous, the atom-based action space of MolDQN produces samples that are heavily biased towards these atoms, resulting in molecules that are well optimized for the task but neither synthetically-accessible nor drug-like. This generation bias was not observable in previously reported benchmarks where atom types were only limited to Carbon, Oxygen, Nitrogen, Sulfur and Halogens \cite{zhou_optimization_2019}. Furthermore, MolDQN samples for the DRD2 task lack a ring system, and whereas molecules from GCPN have one, none adequately optimize for the objective. 

\begin{figure}[!h]
    \centering
    \includegraphics[width=1\linewidth]{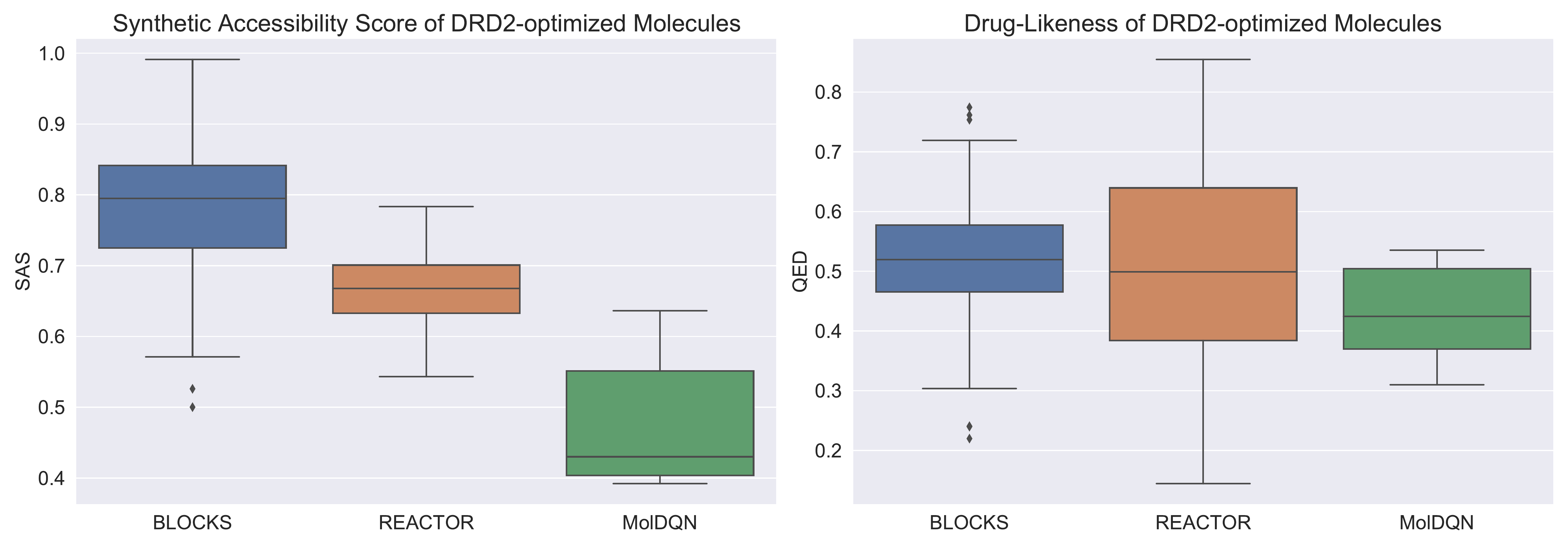}
    \caption{Synthetic accessibility and drug-likeness score distributions of molecules optimized for DRD2 and for the starting blocks.}
    \label{fig:desirability}
\end{figure}

In contrast, REACTOR appears to produce more pharmacologically desirable compounds, without explicitly considering this as an optimization objective. This is supported by Figure \ref{fig:desirability}, which illustrates the shift in synthetic accessibility scores\cite{sas} and drug-likeness for the DRD2-active molecules produced by REACTOR and MolDQN. This suggests that REACTOR is able to simultaneously solve the DRD2 task while maintaining favourable distributions for synthetic-accessibility and drug-likeness.

Further, as shown in Figure~\ref{fig: trajectories} and Figure~\ref{fig: trajectories_2}, optimized compounds are provided along with a possible route of synthesis. While such trajectories may not be optimal, given that they are limited by the reward design and the set of reaction templates available, they provide a crucial indication of synthesizability. Further, it is possible to encourage trajectories to be more efficient by limiting the number of synthesis steps per episode, or by incorporating additional costs such as reactant availability and synthesis difficulty in the reward design. In certain applications, it may also be desirable to increase specificity of the reaction templates via group protection. \citeauthor{Gao_Coley_2020} \cite{Gao_Coley_2020} detail the lack of consideration for synthetic tractability in current molecular optimization approaches, highlighting that this is a necessary requirement for application of these methods in drug discovery. While alternative ideas aiming to embed synthesizability constraints into generative models have recently been explored \cite{chembo,moleculechef,button_automated_2019}, REACTOR is the first approach which explicitly addresses synthetic feasibility by optimizing \textit{directly} in the space of synthesizable compounds using reinforcement learning.

\subsection{Multi-Objective Optimization}

Practical methods for computational drug design must be robust to the optimization of multiple properties. Indeed, beyond the agonistic or antagonistic effects of a small molecule, properties such as the selectivity, solubility, drug-likeness and permeability of a drug candidate must be considered. To validate the REACTOR framework under this setting, we consider the task of optimizing for selective DRD2 ligands. Dopamine receptors are grouped into two classes: D1-like receptors (DRD1, DRD5) and D2-like receptors (DRD2, DRD3 and DRD4). Although these receptors are well studied, design of drugs selective across subtypes remains a considerable challenge. In particular, as DRD1 and DRD3 share 78\% structural similarity in their transmembrane region \cite{dopaminereview,dopaminereceptor}, it is very challenging to identify small molecules that can selectively bind to and modulate their activity. We therefore assess performance both on selectivity across classes (using DRD1 as off-target) and within classes (using DRD3 as off-target). We then analyze how our framework performs as we increase the number of design objectives. For these experiments, we focus our comparison on MolDQN, as it outperforms other state-of-the-art methods on the single-objective tasks. Our approach in combining multiple objectives is that of reward scalarization. Formally, a vector of reward signals is aggregated via a mapping $\mathcal{S}: R_1 \times ... \times R_k \rightarrow \mathbb{R}$, thus collapsing the multi-objective MDP \cite{momdp} into a standard MDP formulation. While the simplest and most common approach to scalarization is to use a weighted sum of the individual reward signals, we adopt a Chebyshev scalarization scheme \cite{chebyshev}, whereby reward signals are aggregated via the weighted Chebyshev metric:

\begin{equation}
    r = - \max_{i} w_i(|r_i - z_i^{*}|) 
\label{eq: cheby}
\end{equation}

where $\Vec{z^*}$ is a utopian vector, $\Vec{w}$ assigns the relative preferences for each objective, and $i$ indexes the objectives. For our experiments, we consider rewards which are constrained to a range between 0 and 1, such that the utopian point is always $\Vec{1}$, rendering the dynamics of each reward signal more comparable, and assign equal preferences to the objectives. For the selectivity tasks, given that both rewards are binary, we use a soft version of this scalarization scheme corresponding the negative euclidean distance to the optimal point. This allows the reward signal to differentiate between reaching 0,1 or both of the objectives. While Chebyshev scalarization was introduced for the setting of tabular Reinforcement Learning, we may interpret it in the function approximation setting as defining an adaptive curriculum, whereby the optimization focus shifts dynamically according to the objective most distant from $\Vec{z^*}$.

\subsubsection{DRD2 Selectivity}
\begin{table*}[h]
    \centering
    \caption{DRD2 Selectivity} 
\resizebox{\textwidth}{!}{%
\begin{tabular}{ccccccc}
\toprule
Objective & Method & Total Actives & Mean Reward & Diversity & Scaff. Similarity & Uniqueness\\
\midrule
D2/D1 
     & MolDQN  & 9.0 $\pm$ 1.41 &   0.64 $\pm$ 0.07 &  0.502 $\pm$ 0.01 & N/A &  0.14 $\pm$ 0.01 \\

     & REACTOR & \textbf{36.667 $\pm$ 4.99} &  0.368 $\pm$ 0.05 & 0.599 $\pm$ 0.01 &  0.139 $\pm$ 0.01 &  0.997 $\pm$ 0.0 \\
\midrule   
D2/D3 
     & MolDQN &   25.667 $\pm$ 3.09 &  0.884 $\pm$ 0.07 &  0.746 $\pm$ 0.05 & N/A &  0.29 $\pm$ 0.02 \\

     & REACTOR & \textbf{53.0 $\pm$ 8.29} & 0.53 $\pm$ 0.08 &  0.692 $\pm$ 0.03 &  0.147 $\pm$ 0.01 &  1.0 $\pm$ 0.0 \\
\bottomrule
\end{tabular}
}
\label{tab: denovo_selectivity}
\end{table*}{}

The total number of actives in Table \ref{tab: denovo_selectivity} corresponds to the number of unique molecules which were found to satisfy all objectives,  while the mean reward in Table \ref{tab: denovo_selectivity} and Figure \ref{fig:moo_ablation} is computed as the proportion of evaluation episodes for which the algorithms optimize all desired objectives. In Table \ref{tab: denovo_selectivity}, we find that REACTOR maintains the ability to identify a higher number of desirable molecules on the selectivity tasks, optimizing for DRD2 while avoiding off-target activity on the D1 and D3 receptors. Further, it is able to outperform MolDQN while maintaining very low scaffold similarity among generated molecules. 

\subsubsection{Robustness to Multiple Objectives}

In addition to off-target selectivity, we assess the robustness of each method's performance as we increase the number of pharmacologically-relevant property objectives to optimize. Specifically, we compare the following combinations of rewards: 

\begin{itemize}
    \item DRD2 with range-targeted cLogP (2 objectives) according to the Ghose filter \cite{ghose}
    \item DRD2, range cLogP, and a molecular weight ranging from 180 to 600 (3 objectives)
    \item DRD2, range cLogP, target molecular weight, and drug absorption, as indicated by a model trained on data for the Caco-2 permeability assay \cite{caco} (4 objectives)

\end{itemize}

For the range-targeted cLogP, molecular weight, and permeability objectives, the component-wise reward is 0 when the molecule falls within the desired range. Otherwise, the distance to the objective is mapped to a range of (0,1]. Given that the DRD2 objective is binary, this implicitly prioritizes the optimization for this reward.

\begin{figure}[!htp]
    \centering
    \includegraphics[width=\linewidth]{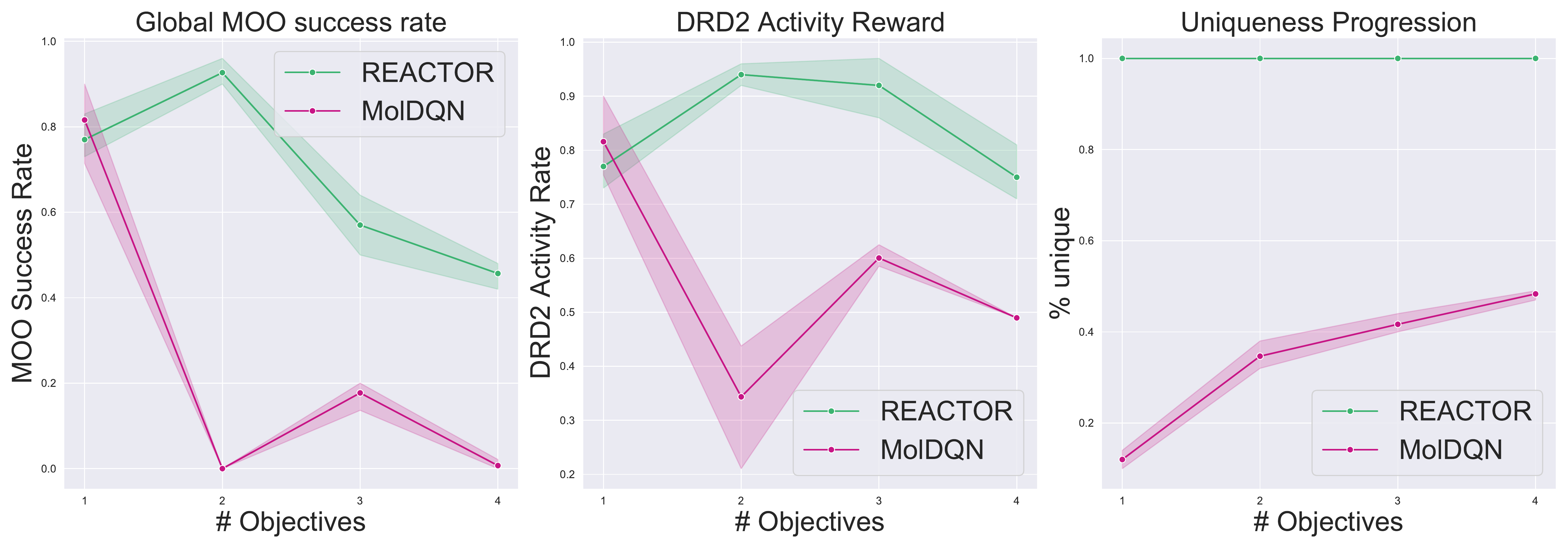}
    \caption{Reward progression as the number of optimization objectives increases.}
    \label{fig:moo_ablation}
\end{figure}

Figure \ref{fig:moo_ablation} shows that REACTOR demonstrates greater robustness to an increasing number of design objectives. Indeed, while both methods see diminishing success rates in optimizing for multiple objectives, the performance of REACTOR diminishes gradually, while MolDQN's performance collapses. Furthermore, REACTOR maintains the ability to generate unique terminal states throughout.

\subsection{Goal-Directed Exploration}
\begin{figure}[!h]
\centering
    \begin{subfigure}{.49\linewidth}
    \centering
    \includegraphics[width=1\linewidth]{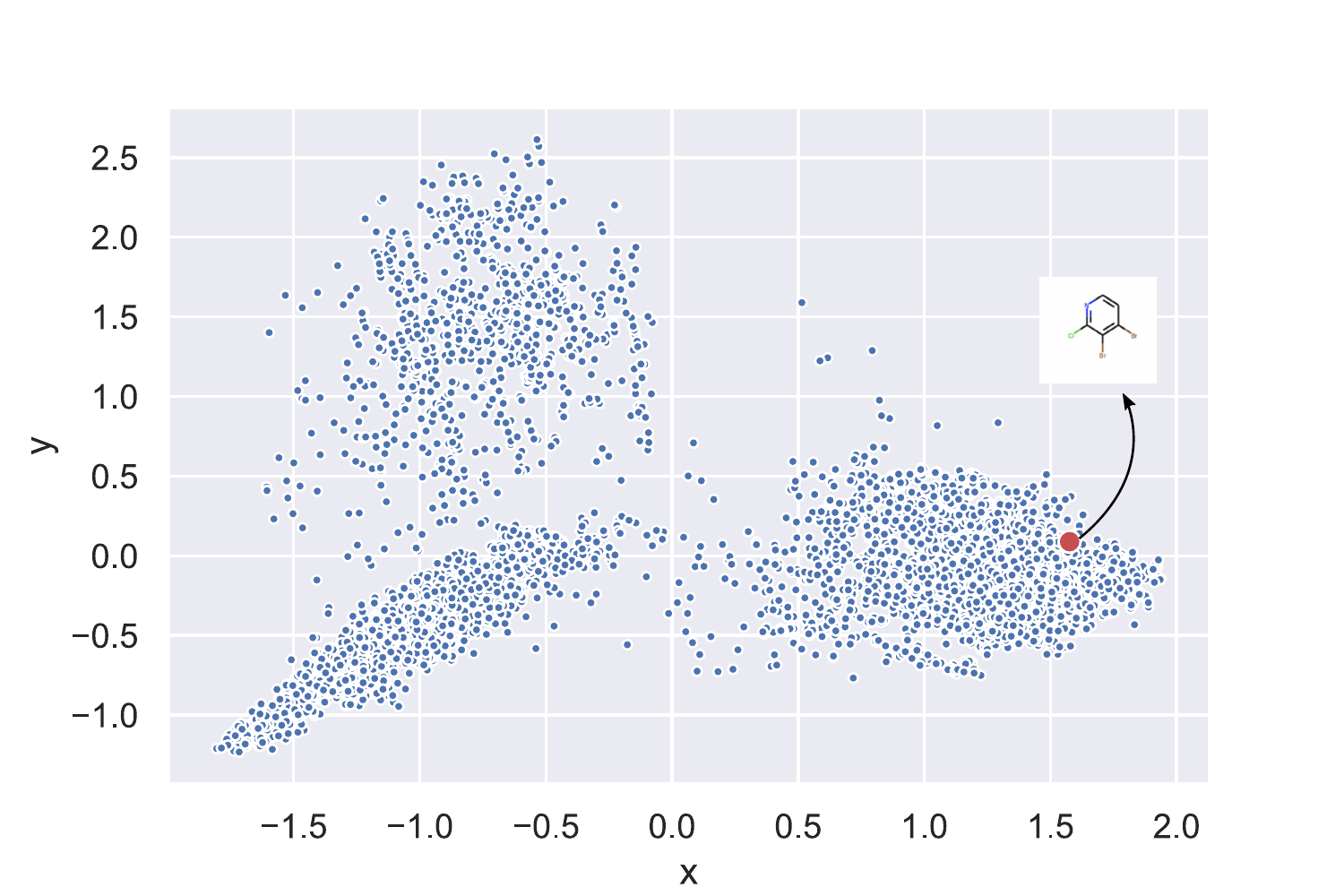}
    \caption{Trajectory Initialization}
    \end{subfigure}%
    \begin{subfigure}{.49\linewidth}
    \centering
    \includegraphics[width=1\linewidth]{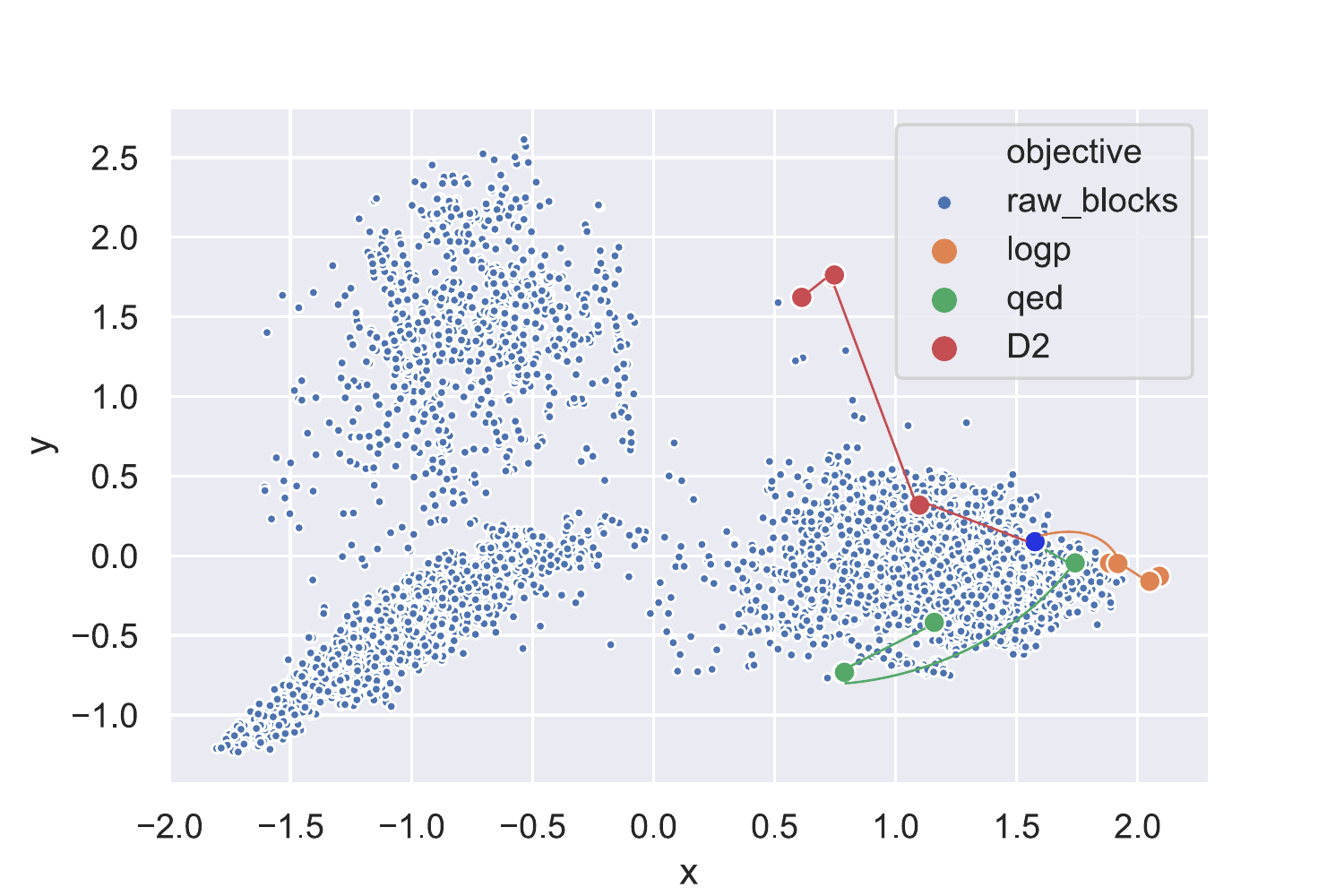}
    \caption{Episode Termination}
    \end{subfigure}
    \caption{Trajectory steps of the REACTOR agent for each objective, starting with the same building block. The RL agent shifts the molecule towards different regions in space to identify the relevant local maximum.}
\label{fig:explore}
\end{figure}

\begin{figure}[!htp]
\centering
    \begin{subfigure}{.49\linewidth}
    \centering
    \includegraphics[width=1\linewidth]{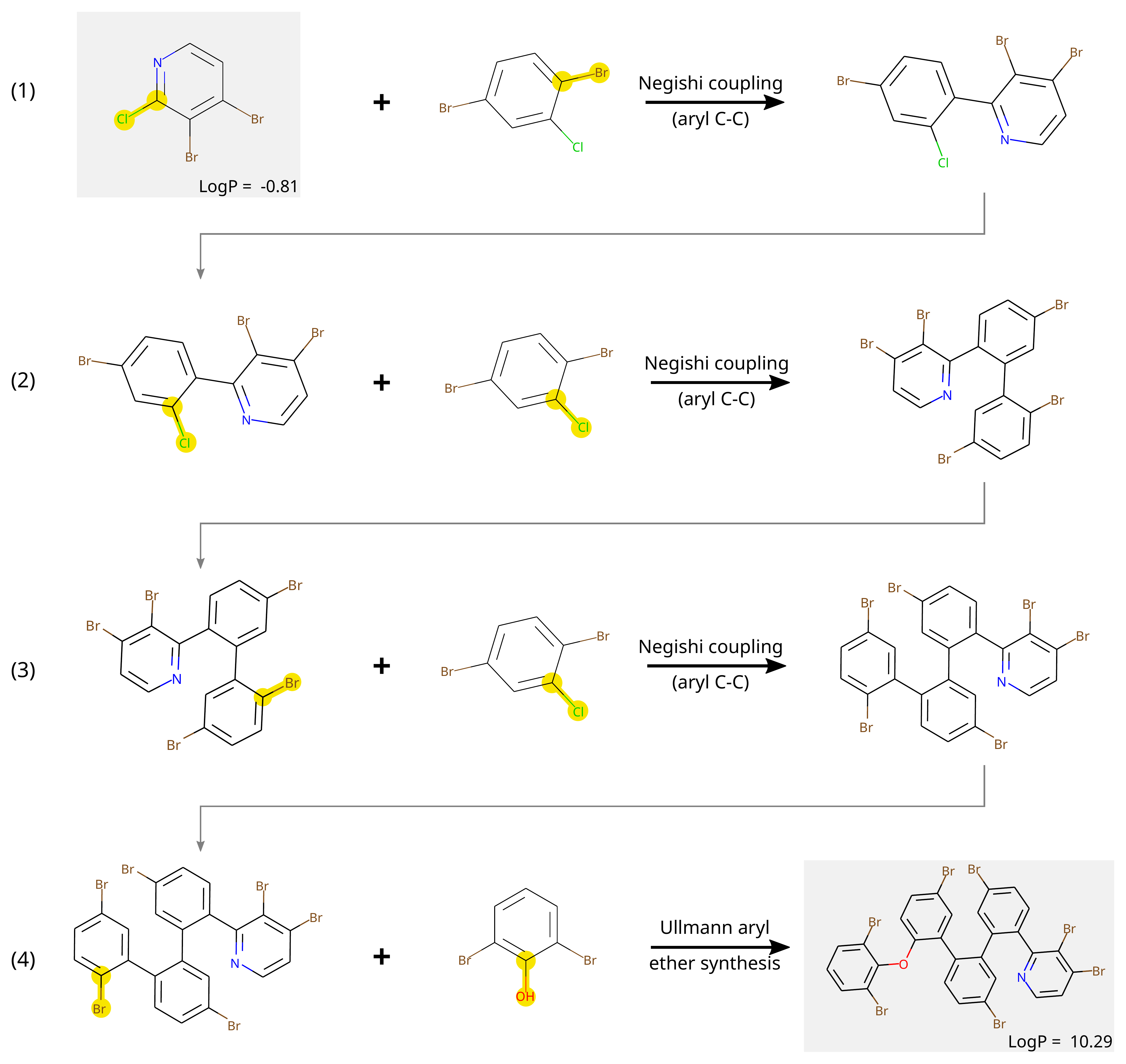}
    \caption{cLogP Trajectory}
    \end{subfigure}
    \begin{subfigure}{.49\linewidth}
    \centering
    \includegraphics[width=1\linewidth]{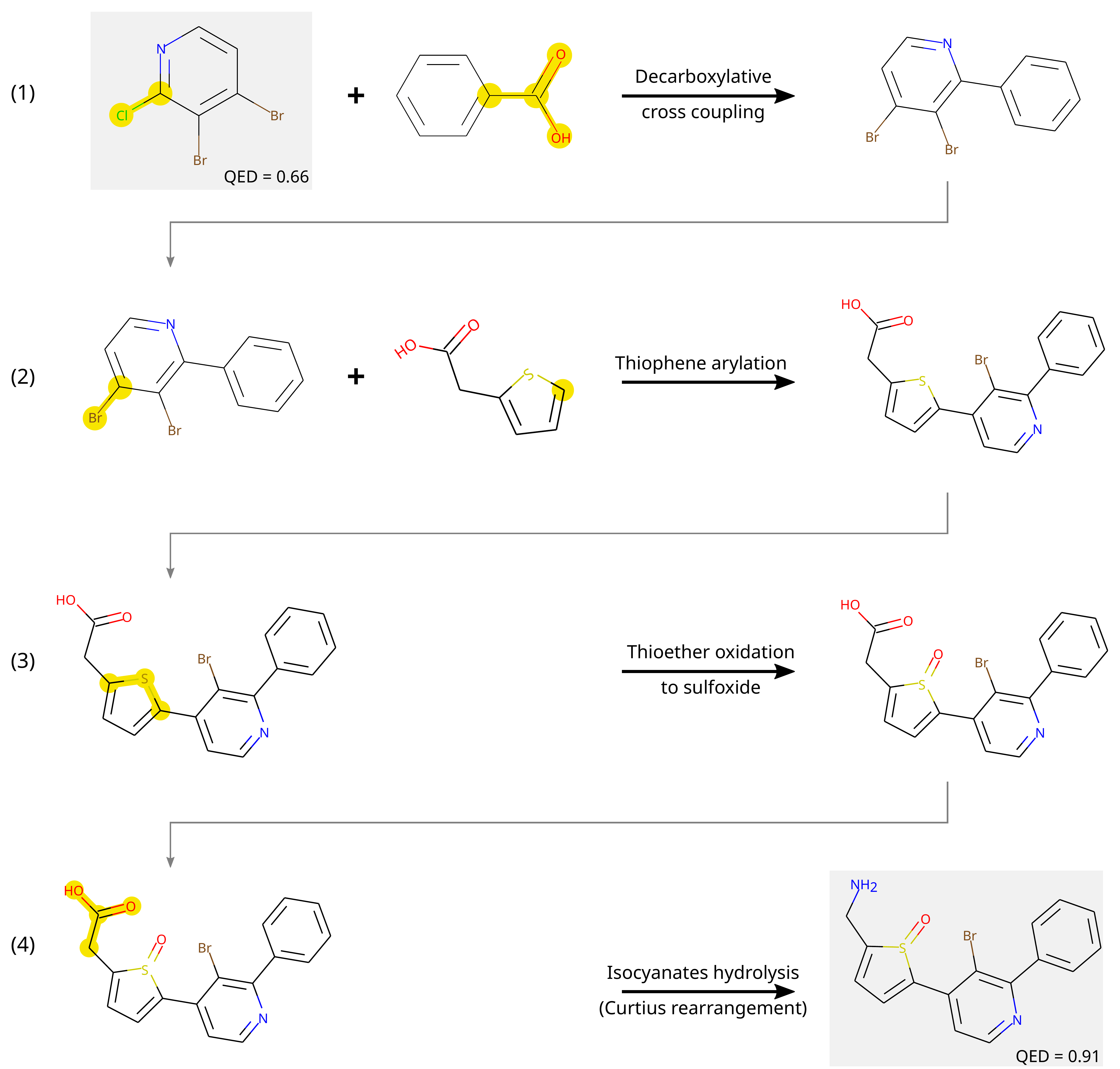}
    \caption{QED Trajectory}
    \end{subfigure}
    \caption{Trajectories taken by the REACTOR agent from the same building block for different objectives. Note that the reaction steps are simplified and are mainly indicative of synthesizability. For example, the Negishi coupling reaction would first require the formation of an organozinc precursor. Furthermore, selectivity is low at some steps, which will result in a mixture of products, unless reacting groups are protected.}
    \label{fig: trajectories_2}
\end{figure}

In order to gain further insight into the nature of the trajectories generated by the REACTOR agent, we plotted two alternative views of optimization routes generated for the same building block across each single-property objective. In Figure \ref{fig:explore}, we fit a Principal Components Analysis (PCA) \cite{pca} on the space of building blocks to identify the location of the initial state, and subsequently transform the next states generated by the RL agent onto this space. We find that the initial molecule is clearly shifted to distinct regions in space, while the magnitude of the transitions suggest efficient traversal of the space. This provides further evidence that exploration through space is a function of reward design, and is mostly unbiased by the data distribution of initialization states.
Figure \ref{fig: trajectories_2} shows the same trajectories with their corresponding reactions and intermediate molecular states. We find that optimized molecules generally contain the starting structure. We believe this to be a desirable property given that real-life design cycles are often focused on a fixed scaffold or set of core structures. We also note that the policy learned by our REACTOR framework is able to generalize over different starting blocks, suggesting that it achieves generation of structurally diverse and novel compounds.

{}

\section{Conclusion}
\label{sec: conclusion}
This work proposes a novel approach to molecular design which defines state transitions as chemical reactions within a reinforcement learning framework. We demonstrate that our framework leads to globally improved performance, as measured by reward and diversity of generated molecules, as well as greater training efficiency while producing more drug-like molecules. Analysis of REACTOR's robustness to multiple optimization criteria, coupled with its ability to maintain predicted activity on the DRD2 receptor, suggests increased potential for successful application in drug discovery. Furthermore, molecules generated by this framework exhibit better synthetic accessibility by design, with one viable synthesis route also suggested. Although the reactivity and stability of the optimized molecules remain a potential issue, REACTOR's efficiency in a multiple optimization setting suggests that this can be addressed by explicitly considering them as additional design objectives.

Future work aims to build on this framework by making use of its hierarchical formulation to guide agent policies both at the higher reaction and lower reactant levels, exploring proposals from h-DQN \cite{hdqn} for hierarchical value functions, or the option-critic framework \cite{Bacon_Harb_Precup_2016} as a starting point. We also plan to expand the effective state space of our MDP by embedding a synthesis model, with Transformer-based architectures showing promise \cite{transformer}, as the MDP transition model. Because practical de novo design requires optimization of multiple criteria simultaneously, we believe the efficiency of our design framework provides a robust foundation for such tasks, and hope to expand on existing approaches \cite{abels_dynamic_2019, moffaert_multi-objective_2014, yang_generalized_2019} for multi-objective reinforcement learning. Finally, we intend to validate the proposed synthetic routes and bio-activity of generated molecules experimentally to better demonstrate real-world utility.

\subsubsection*{Acknowledgments} \label{sec: Acknowledgment}
The authors thank Daniel Cohen, Sébastien Giguère, Violet Guo, Michael Craig, Connor Coley and Grant Wishart for reviewing the manuscript and for their helpful comments. The methods and algorithms presented here were developed at InVivo AI.


\newpage
\small
\bibliography{biblio}\newpage

\renewcommand{\thepage}{S\arabic{page}}
\setcounter{page}{1}
\section{Supplementary Material}
\appendix
\beginsupplement

\section{Results}

\begin{table*}[h]
	\centering
	\caption{Goal-Directed Molecule Design} \label{tab:benchmark}
	\resizebox{\textwidth}{!}{%
		\begin{tabular}{ccccccc}
			\toprule
			Objective & Method & Mean Reward & Max Reward & Diversity & Scaff. Similarity & Uniqueness\\
			\midrule
			cLogP & BLOCKS & -1.80 $\pm$ 0.08 & 1.80 $\pm$ 0.32 & 0.94 $\pm$ 0 & N/A  &  100\% $\pm$ 0  \\
			\midrule
			& Hill Climbing  & 7.14 $\pm$ 0.20 & 10.90 $\pm$ 0.04 & 0.73 $\pm$ 0.01 & \textbf{0.13 $\pm$ 0.01} & \textbf{100\% $\pm$ 0} \\
			
			& ORGAN  & -2.47 & 0.97 & 0.83 & 0.14 & 63\% \\
			
			& JTVAE  & -1.48 $\pm$ 0.56 & 0.16 $\pm$ 0.14 & 0.54 $\pm$ 0.2 & N/A & 41\% $\pm$ 34\%  \\
			
			& GCPN  & 1.03 $\pm$ 0.28 & 8.51 $\pm$ 0.35 & \textbf{0.90 $\pm$ 0} & 0.20 $\pm$ 0.01 & \textbf{100\% $\pm$ 0} \\
			
			& MolDQN  & \textbf{12.84 $\pm$ 0.23} & \textbf{18.42 $\pm$ 0.37} & 0.71 $\pm$ 0.01 & 0.72 $\pm$ 0.23 & 72\% $\pm$ 3.6\% \\
			
			& REACTOR & 8.01 $\pm$ 0.18 & 10.74 $\pm$ 0.28  &  0.69 $\pm$ 0.01 & 0.20 $\pm$ 0 & 99.7\% $\pm$ 0.5\% \\
			
			\midrule
			QED  & BLOCKS & 0.523 $\pm$ 0.005 & 0.763 $\pm$ 0.009 & 0.94 $\pm$ 0 & N/A  &  100\% $\pm$ 0  \\
			\midrule
			& Hill Climbing & 0.811 $\pm$ 0.007 & 0.943 $\pm$ 0.004 & 0.879 $\pm$ 0.003 & 0.20 $\pm$ 0.023 & \textbf{100\% $\pm$ 0}  \\
			
			& ORGAN  & 0.608 & 0.906 & 0.871 & 0.178 & 89.5\% \\
			
			& JTVAE  & 0.604 $\pm$ 0.017 & 0.876 $\pm$ 0.048 & 0.841 $\pm$ 0.018 & 0.638 $\pm$ 0.046 & 92.8\% $\pm$ 5.5\%  \\
			
			& GCPN  & 0.607 $\pm$ 0.012 & 0.916 $\pm$ 0.012 & \textbf{0.91 $\pm$ 0.002} & \textbf{0.112 $\pm$ 0.004} & \textbf{100\% $\pm$ 0} \\
			
			& MolDQN  & 0.857 $\pm$ 0.026 & 0.936 $\pm$ 0.004 & 0.791 $\pm$ 0.007 & 0.620 $\pm$ 0.123 & 67\% $\pm$ 5.8\% \\
			
			& REACTOR & \textbf{0.876 $\pm$ 0.007} & \textbf{0.947 $\pm$ 0.001} & 0.878 $\pm$ 0.002 & 0.161 $\pm$ 0.021 & \textbf{100\% $\pm$ 0}\\ 
			
			\bottomrule
		\end{tabular}
	}
	\label{tab: denovo_logp}
\end{table*}{}

\FloatBarrier
\section{Figures}

\begin{figure}[!h]
\centering
    \centering
    \includegraphics[width=1\textwidth]{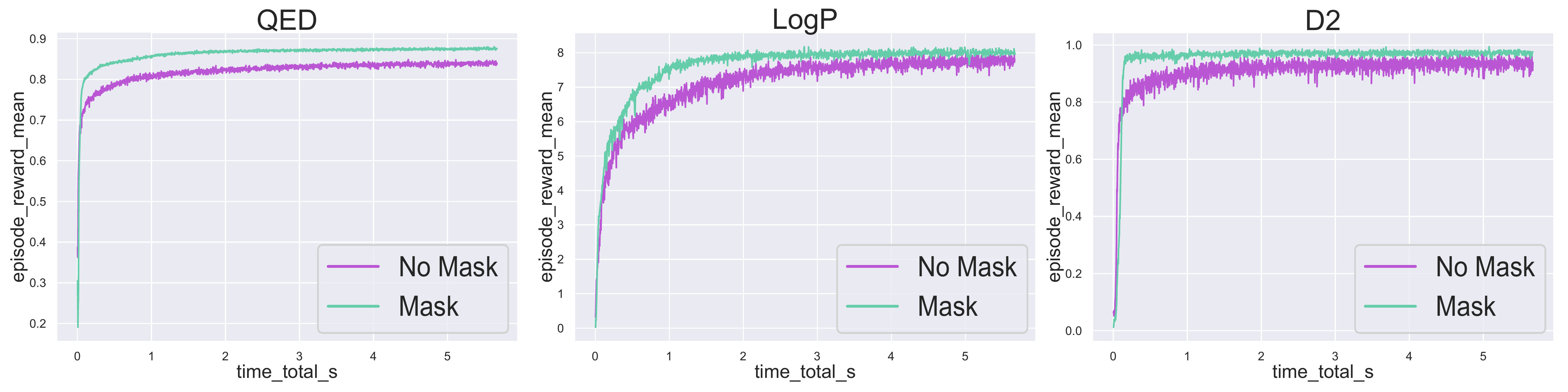}
    \caption{REACTOR convergence ablation when using a masked action space}
\label{fig:masking}
\end{figure}

\begin{figure}[!h]
\centering
    \begin{subfigure}{0.8\linewidth}
    \centering
    \includegraphics[width=0.6\linewidth]{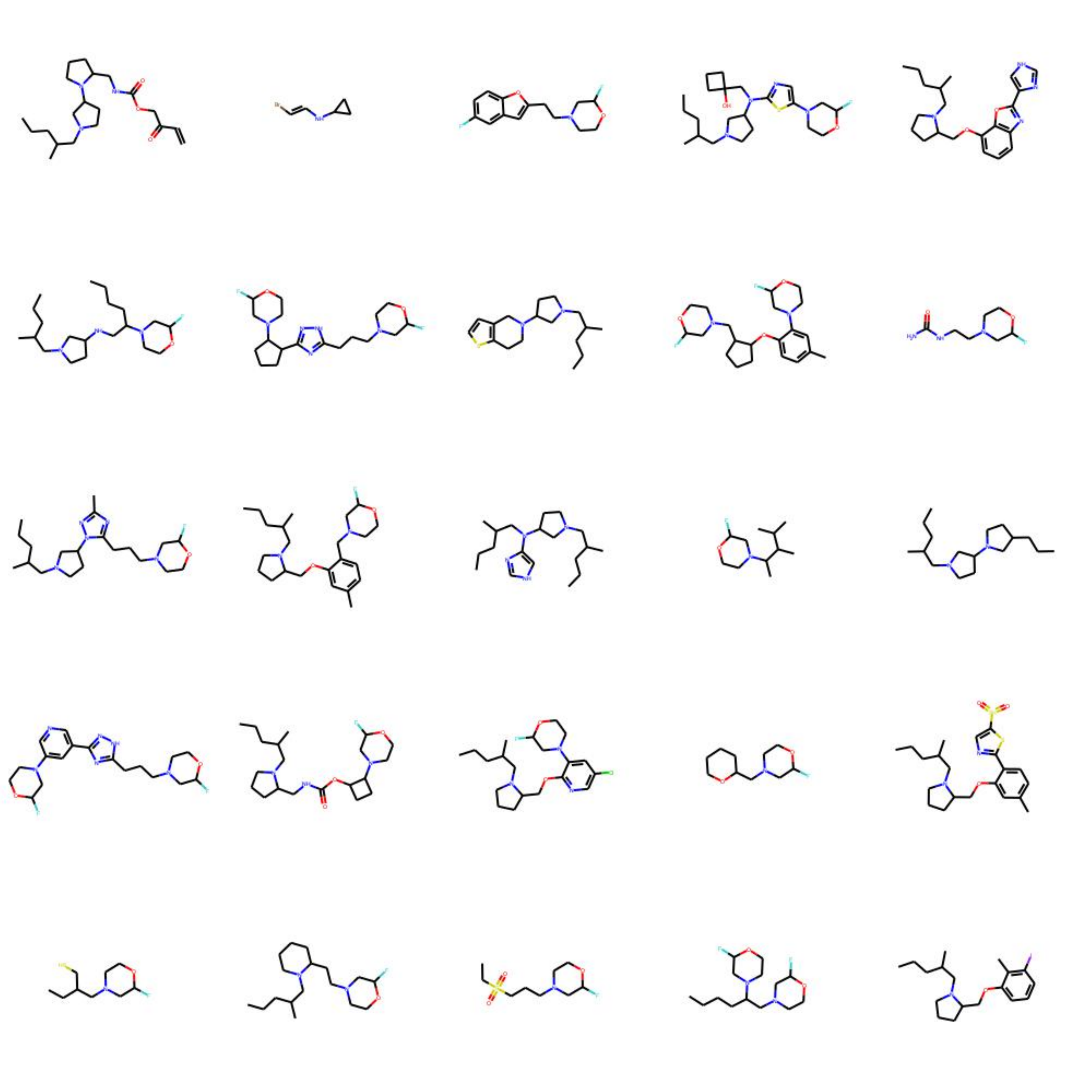}
    \caption{DRD2 Molecule Samples}
    \end{subfigure}
    \begin{subfigure}{0.42\linewidth}
    \centering
    \includegraphics[width=1\linewidth]{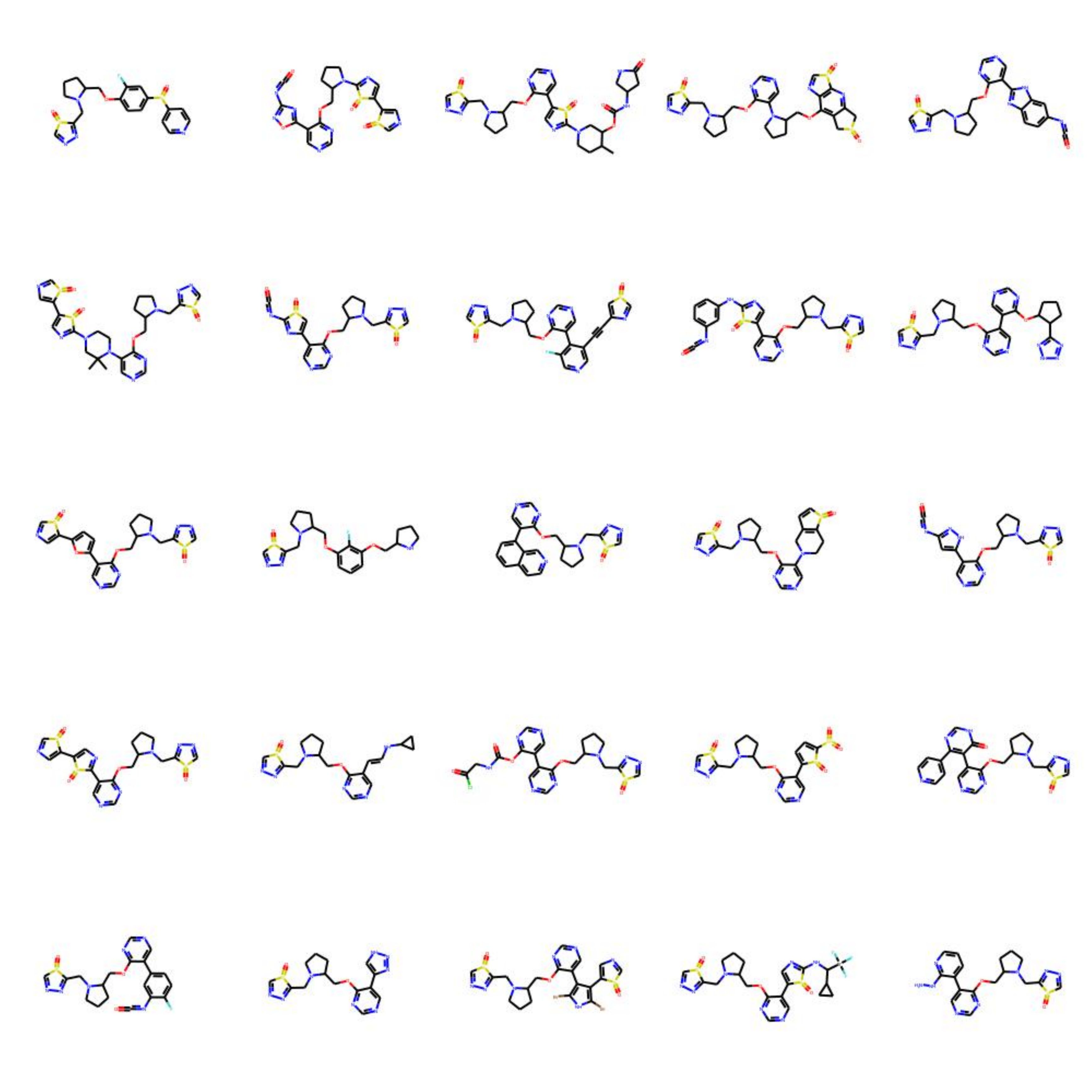}
    \caption{DRD2 with D1 selectivity}
    \end{subfigure}
    \begin{subfigure}{0.42\linewidth}
    \centering
    \includegraphics[width=1\linewidth]{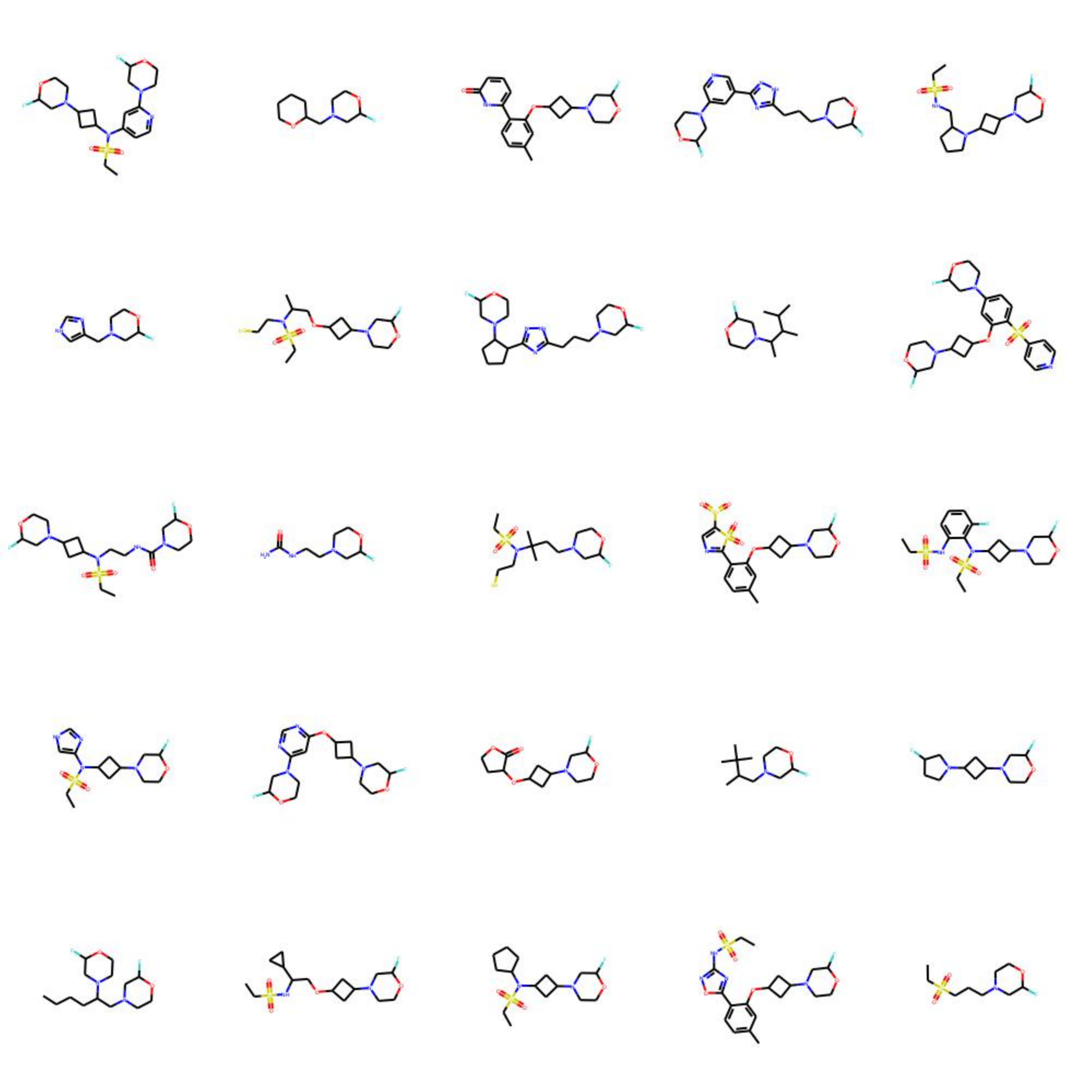}
    \caption{DRD2 with D3 selectivity}
    \end{subfigure}
    \caption{Molecule samples for the various DRD2 optimization tasks}
\label{fig:reaction_samples}
\end{figure}

\FloatBarrier
\section{Reward Model Details}
\subsection{DRD2 Reward Model}
\begin{figure}[!h]
\centering
    \includegraphics[width=\textwidth]{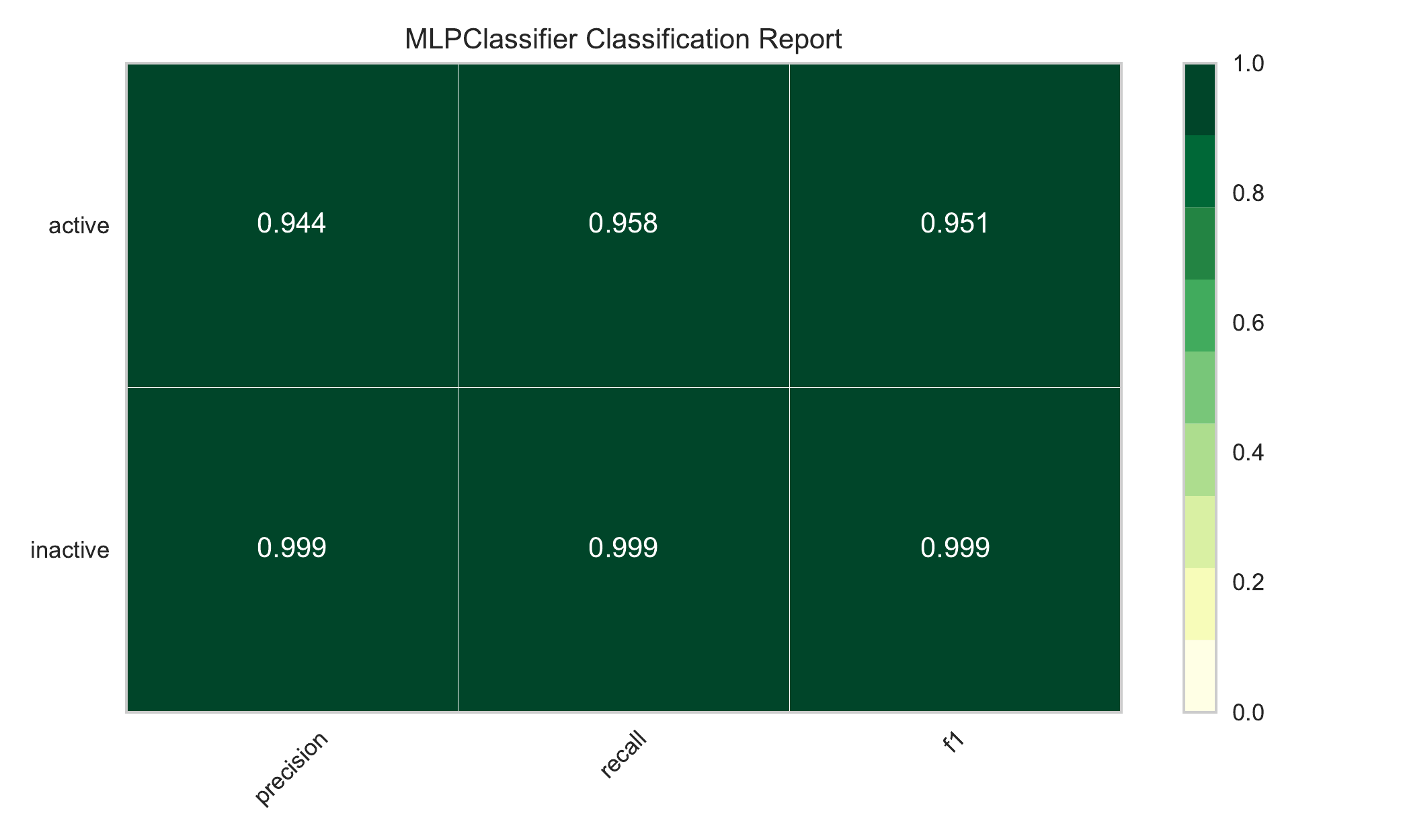}
    \caption{Model performance on test data for the selected DRD2 model}
\label{fig:drd2_cm}
\end{figure}

The model for the DRD2 receptor was trained using data from ExCAPE-DB \cite{excapedb}, with 8323 active and 343206 inactive compounds. Molecules were then sanitized and duplicate molecules were removed. We then performed a stratified split consisting of 90\% training and 10\% test splits. 3-fold cross validation was performed over the training set in order to select a model. We compared Random Forest, Gradient Boosting, Support Vector Machines and Feed-Forward Neural Networks, using 2048 Morgan Fingerprints with radius 2 as molecular featurizations. The selected model is a 200 neuron single-layer neural network, with its classification performance on the held-out test set provided in Figure \ref{fig:drd2_cm}. 

\subsection{DRD1 and DRD3 Reward Models}

DRD1 and DRD3 modulators were obtained from ExCAPEDB \cite{excapedb}. Due to the high data imbalance, only structurally diverse inactive (pXC50 < 5) compounds with experimentally validated activity were retained. Each dataset was subsequently cleaned using the following procedure:

\begin{itemize}
    \item All molecules are sanitized and standardized.
    \item Duplicate compounds were removed and only the largest fragment was retained for each molecule. This resulted in a dataset of 1753 actives vs 10317 inactives for DRD1 and 3498 vs 10074 inactives for DRD3. Each dataset was split into an 80\% training and 20\% test set, using a stratified split.
    \item The DRD1 and DRD3 activity models were trained using cross-validation (80-20) under various splits of the training set (random split, stratified activity split, structural-similarity based clustering split, scaffold split) and evaluated using balanced accuracy and f1-score. We considered various featurizations and their combinations, as well as several machine learning algorithms (Support Vector Machine, Random Forest, Gradient Boosting, Logistic Regression and a Multi-Layer Perceptron). The hyper-parameters, including  molecular featurization, resulting in the best performances were selected for each algorithm, and the best performing model on the held out test set was retained.
\end{itemize}{}

For both datasets, the best model according to the F1-score/ROC-AUC/Balanced Accuracy was a Gradient Boosting Classifier.

\subsection{Caco-2 Reward Model}
Data for the Caco-2 cell permeability assay was obtained from \citeauthor{caco2}, with a measurement unit of $log(10^{-6}) cm/s$. Model selection was performed using a 6-fold stratified split. Algorithms compared at this stage were Random Forest, Kernel Ridge, and Gaussian Process regression algorithms, with model selection additionally performed over various Fingerprint featurizations. The final model is a Kernel Ridge Regression model with a Laplacian kernel, with 512-bit Estate Fingerprints.

\end{document}